\documentclass[reprint,amsmath,amssymb,aps,longbibliography,pra]{revtex4-1}
\usepackage[utf8]{inputenc}
\usepackage{graphicx}
\usepackage{dcolumn}
\usepackage{mathrsfs,amsmath}
\usepackage{bm}
\usepackage{xcolor}
\begin{document}
\title{Muonium hydride: the lowest density  crystal}
\author{Youssef Kora and Massimo Boninsegni$^\star$}
\affiliation{Department of Physics, University of Alberta,  Edmonton, AB T6G 2E1, Canada}
\author{Dam Thanh Son}
\affiliation{Kadanoff Center for Theoretical Physics, The University of Chicago, Chicago, IL 60637, USA}
\author{Shiwei Zhang}
\affiliation{Center for Computational Quantum Physics, Flatiron Institute, New York, NY 10010, USA}
\affiliation{Department of Physics, The College of William and Mary, Williamsburg, VA 23187, USA}
\email{m.boninsegni@ualberta.ca}



\begin{abstract}
A muonium hydride molecule is a bound state of muonium and hydrogen atoms. It has
half the mass of a parahydrogen molecule and very similar electronic properties in its ground state. The phase diagram of an assembly of such particles is investigated by first principle quantum simulations. 
In the bulk limit, the low-temperature equilibrium phase is a crystal of extraordinarily low density, lower than that of any other known atomic or molecular crystal. Despite the low density and particle mass, the melting temperature is surprisingly high (close to 9 K). No (metastable) supersolid phase is observed. 
We investigated the physical properties of nanoscale clusters (up to 200 particles) of muonium hydride and found the superfluid response to be greatly enhanced compared to that of parahydrogen clusters. The possible experimental realization of these systems is discussed.
\end{abstract}

\flushbottom
\maketitle
%
%
\thispagestyle{empty}

\section{Introduction}

An intriguing open question in condensed matter physics, one with potential practical significance, is whether there exists a lower bound for the density of a crystal. In other words, how large can the average distance between nearest neighbor atoms be, before crystalline long-range order is suppressed by thermal and quantum fluctuations? It remains unclear whether a lower limit exists, or 
how to approach it experimentally. We begin our discussion by rehashing a few basic facts about the crystalline phase of matter.
\\ \indent
Crystallization occurs at low temperature ($T$) in almost all known substances, as the state of lowest energy (ground state) is approached. Classically, the ground state is one 
in which the potential energy of interaction among the constituent particles is minimized, 
a condition  that corresponds to an orderly arrangement of particles in regular, periodic lattices. 
In most cases, quantum mechanics affects this fundamental conclusion only quantitatively, as zero-point motion of particles results in lower equilibrium densities and melting temperatures, with respect to what one would predict classically; typically, these corrections are relatively small. Only in helium is the classical picture upended by quantum mechanics, as the fluid resists crystallization all the way down to  temperature $T=0$,K, under the pressure of its own vapor.
\\ \indent
In many 
condensed matter systems, the interaction among atoms 
is dominated by pairwise contributions, whose ubiquitous features are a strong repulsion at interparticle distances less than a characteristic length $\sigma$, as well as an attractive part, which can be described by an effective energy well depth $\epsilon$. The dimensionless parameter $\Lambda=\hbar^2/[m\epsilon\sigma^2]$, where $m$ is the particle mass, quantifies the relative importance of quantum-mechanical effects in the ground state of the system. \\ \indent
It has been established  for a specific model pair potential incorporating the above basic features, namely the 
Lennard-Jones potential,
and for Bose statistics, 
that the equilibrium phase is a crystal if $\Lambda < \Lambda_c \approx 0.15$; it is a (super)fluid if $0.15\lesssim \Lambda\lesssim 0.46$, while, for $\Lambda > 0.46$, the system only exists in the gas phase \cite{pnas}. The value of $\Lambda_c$ is estimated \cite{nosanow} to be slightly higher ($\approx$ 20 \%) for systems obeying Fermi statistics. For substances whose elementary constituents are relatively simple atoms or molecules, this result provides a useful, general criterion to assess their propensity to crystallize, as measured by their proximity in parameter space to a quantum phase transition to a fluid phase. It can be used to infer, at least semi-quantitatively, macroscopic properties of the crystal, such as its density and melting temperature, both of which generally decrease \cite{pnas} as $\Lambda\to\Lambda_c$. 
\\ \indent
The two stable isotopes of helium have the highest known values of $\Lambda$, namely 0.24 (0.18) for $^3$He ($^4$He); for all other 
naturally occurring substances, $\Lambda$ is considerably lower. The next highest value is that of molecular hydrogen (H$_2$), namely $\Lambda = 0.08$, quickly decreasing for heavier elements and compounds. At low temperature, H$_2$ forms one of the least dense crystals known, of density $\rho=0.0261$ \AA$^{-3}$ (mass density 0.086 gr/cc). The low mass of a H$_2$ molecule (half of that of a $^4$He atom) and its bosonic character (its total spin is $S=0$ in its ground state) 
led to the speculation that liquid H$_2$ may turn superfluid at low temperature \cite{ginzburg}, just like $^4$He. In practice, no superfluid phase is observed (not even a metastable one), as molecular interactions, and specifically \cite{boninsegni18} the relatively large value of $\sigma$ ($\sim 3$ \AA) cause H$_2$ to crystallize at a temperature $T=13.8$\,K.
\\ \indent
Present consensus is 
that H$_2$ is a non-superfluid, insulating crystal at low temperature, including in reduced dimensions \cite{2d,1d}; only small clusters of parahydrogen ($\sim 30$ molecules or less) are predicted \cite{sindzingre,mezz1,mezz2,2020} to turn superfluid at $T\sim 1$ K, 
for which some experimental evidence has been reported \cite{grebenev}.
If, hypothetically, the mass of the molecules could be progressively reduced, while leaving the interaction unchanged, thus increasing $\Lambda$ from its value for H$_2$ all the way to $\Lambda_c$, several intriguing scenarios might arise, including a low temperature superfluid liquid phase, freezing into a low-density crystal at $T=0$, and even a supersolid phase, namely one enjoying at the same time crystalline order and superfluidity \cite{supersolid}. Obviously, the value of $\Lambda$ can also be modified by changing one or both of the interaction parameters ($\epsilon$ and $\sigma$) independently of the mass; in this work, however, we focus for simplicity on the effect of mass on the physics of the system.
\\ \indent
One potential way to tune the mass is via substitution of protons or 
electrons with muons. For example, an 
assembly of molecules of muonium hydride (HMu) 
differ from H$_2$ by the replacement of one of the two protons with an antimuon ($\mu^+$), whose mass is approximately 11\% of that of a proton. 
Quantum chemistry calculations 
have shown that it is very similar in size to H$_2$, and has the same quantum numbers in the ground state
\footnote{Specifically, the average distance between the proton and the muon in a HMu molecule is estimated at 0.8 \AA, whereas that between the two protons in a H$_2$ molecule is 0.74 \AA. See Refs, \cite{suff,zhou}}.
It is therefore not inconceivable that the interaction between two HMu might be quantitatively close to that between two H$_2$ 
(we further discuss this aspect below). In this case the value of the parameter $\Lambda$ 
is $\sim 0.14$, i.e., very close to $\Lambda_c$ for a Bose system. This leads to the speculation that this substance may crystallize into a highly quantal solid, displaying a strikingly unique behavior, compared to ordinary crystals.
In order to investigate such a scenario, and more specifically to gain insight into the effect of mass reduction, we studied  theoretically the low temperature phase diagram of a hypothetical condensed phase of HMu,
by means of first principle computer simulations of a microscopic model derived from that of H$_2$. 
\\ \indent
The main result is that the equilibrium phase of the system at low temperature is a crystal of very 
low density, some $\sim 5\%$ lower than the $T=0$ equilibrium density of {\em liquid} $^4$He. Despite the low density, however, such a crystal melts at a fairly high temperature, close to 9 K, i.e., only a few K lower than that of H$_2$. No superfluid phases are observed, either fluid or crystalline, as exchanges of indistinguishable particles, known to underlie superfluidity \cite{feynman}, are strongly suppressed in this system, much as they are in H$_2$, by the relatively large size of the hard core repulsion at short distances (i.e., $\sigma$). As a result, the behavior of this hypothetical system can be largely understood along classical lines.
This underscores once again the crucial role of exchanges of identical particles in destabilizing the classical picture, which is not qualitatively altered by zero-point motion of particles alone \cite{role}; it also reinforces the conclusion \cite{boninsegni18} reached elsewhere that it is the size of the hard core diameter of the intermolecular interaction that prevents superfluidity in H$_2$, {\em not} the depth $\epsilon$ of its attractive part.
\\ \indent
To complement our investigation, we also studied nanoscale HMu 
clusters of varying sizes, comprising up to few hundred molecules. We find the behavior of these clusters to be much closer to that of $^4$He (rather than H$_2$) clusters. For example, at $T=1$ K the structure of HMu clusters is liquid-like, and their superfluid response  approaches 100\%, even for the largest cluster considered (200 HMu molecules). Thus, while mass reduction does not bring about substantial physical differences between the behavior of bulk HMu and that of H$_2$, it significantly differentiate the physics of nanoscale size clusters.
\section{Model}\label{mod}
\begin{figure}[ht]
\centering
\includegraphics[width=\linewidth]{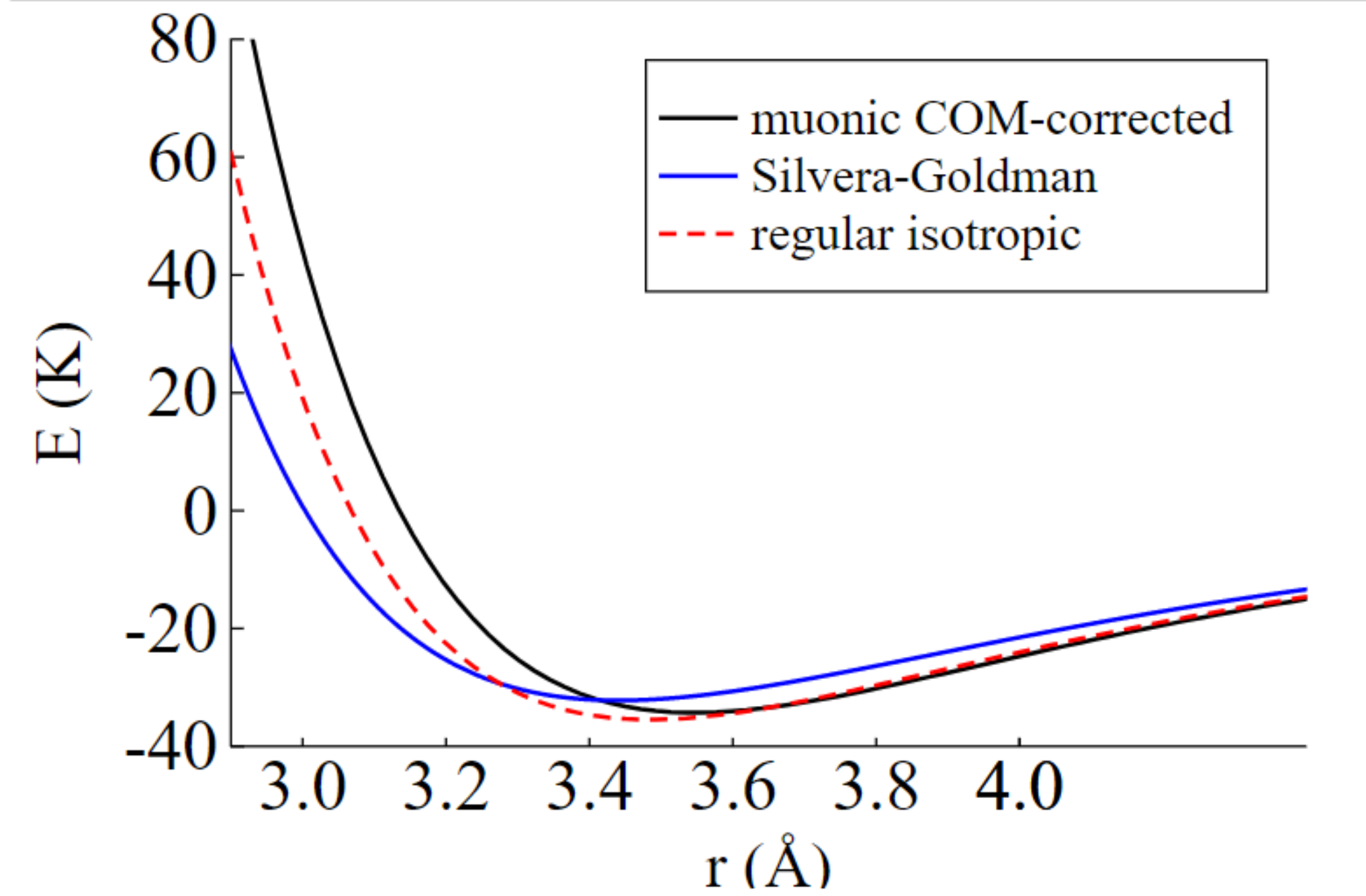}
\caption{The intermolecular interaction energy $E$ (in K) as a function of distance (\AA) between the centers of mass of the molecules, for 
the  H$_2$ (red, dashed) and  HMu 
(black, solid line) cases, obtained with the aid of the programs provided in Ref. \cite{Szal}. Also shown (blue, solid line) is the Silvera-Goldman potential.}
\label{fig:pot}
\end{figure}

We model the HMu molecules as point-like, identical particles of mass $m$ and spin $S=0$, thus obeying Bose statistics.
For the bulk studies, the system is enclosed in a cubic cell of volume $V=L^3$ with periodic boundary conditions in the three directions, giving a
density of $\rho=N/V$. 
For the cluster studies, the $N$ particles  are 
placed in a supercell of large enough size to remove boundary effects.
The quantum-mechanical many-body Hamiltonian reads as follows:
\begin{eqnarray}\label{u}
\hat H = - \lambda \sum_{i}\nabla^2_{i}+\sum_{i<j}v(r_{ij})
\end{eqnarray}
where the first (second) sum runs over all particles (pairs of particles), $\lambda\equiv\hbar^2/2m=21.63$ K\AA$^{2}$ (reflecting the replacement of a proton with a $\mu^+$ in a H$_2$ molecule), $r_{ij}\equiv |{\bf r}_i-{\bf r}_j|$ and $v(r)$ denotes the pairwise interaction between two HMu molecules, which is assumed spherically symmetric. 
\begin{figure}[h]
\centering
\includegraphics[width=2.0in]{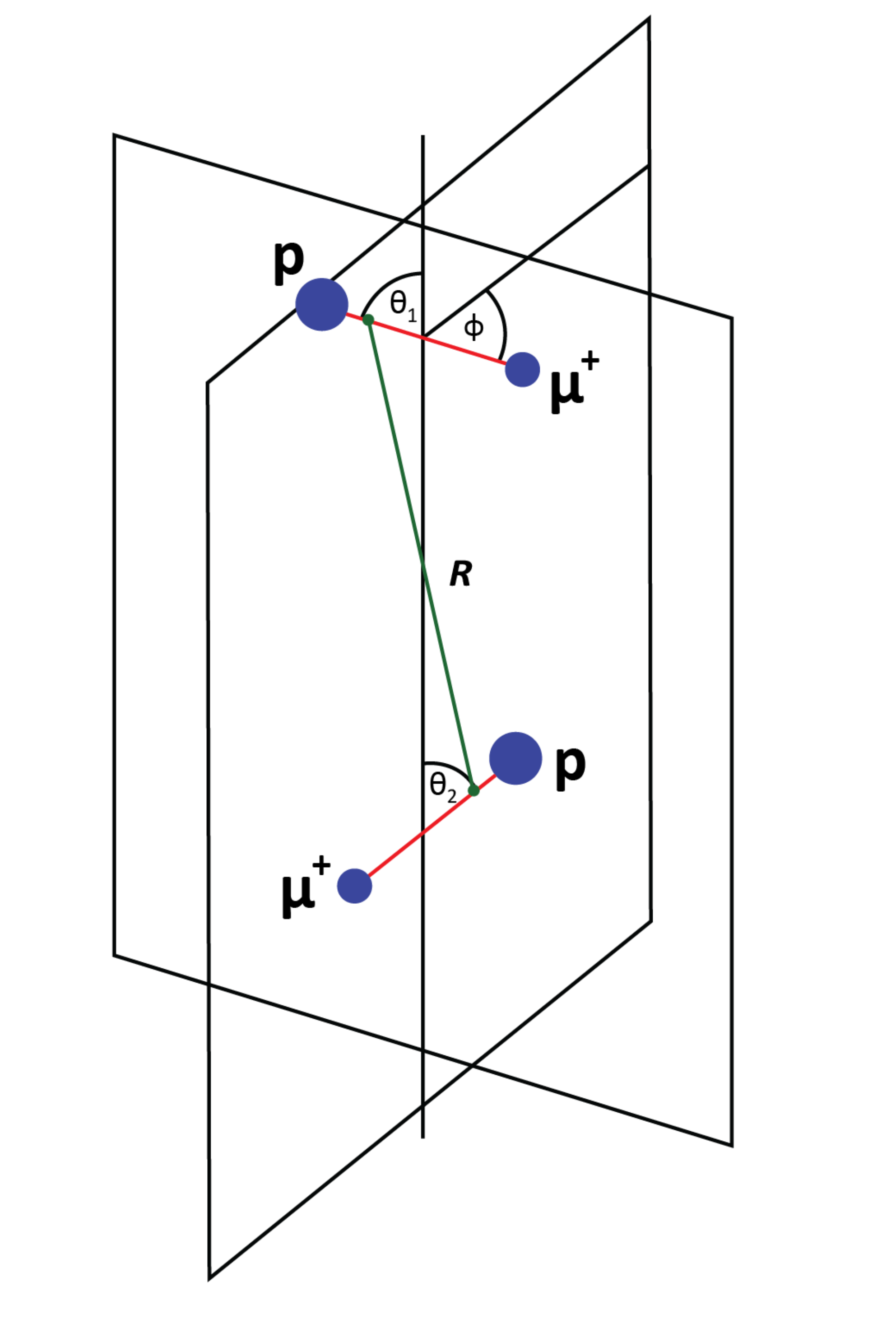}
\caption{Geometry utilized in the calculation of the effective interaction between two HMu molecules. Shown is the line connecting the centers of mass of the two molecules, as well the three angles describing their relative orientation, i.e., the two polar angles $\theta_1, \theta_2$, as well as the azimuthal angle $\phi$.}
\label{geom}
\end{figure}
\\ \indent
In order to decide on an adequate model potential to adopt in our calculation, we use as our starting point the H$_2$  intermolecular potential, for which a considerable amount of theoretical work has been carried out \cite{SG,DJ,Szal}. We consider here for definiteness the {\em ab initio} pair potential proposed in  Ref. \onlinecite{Szal}.
The 
most important effect of the replacement of one of the protons of the H$_2$ molecule with a $\mu^+$ is the shift of the center of mass of the molecule away from the midpoint, and toward the proton. We assume that this effect provides the leading order correction with respect to the H$_2$ intermolecular potential, and we set out to obtain a corrected version of the interaction for the new geometry, as illustrated in Fig.~\ref{geom}. We use the program provided in Ref. \cite{Szal} to generate a potential as a function of the distance between the midpoints and the angular configurations, 
and then transform that to a potential as a function of the distance between the centers of mass and the angular configurations. Finally, we average over the angular configurations to obtain a one-dimensional isotropic potential. We take the distance between the proton and the $\mu^+$ to be that computed in Ref. \cite{zhou}, which differs only slightly from the distance between the two protons in the H$_2$ molecule.
\\ \indent
In Fig. \ref{fig:pot}, we compare the potential energy of interaction between two HMu molecules resulting from this procedure with that of obtained for two H$_2$ molecules, i.e., with the center of mass at the midpoint. The comparison suggests that the displacement of the center of mass of the molecule results in a slight stiffening of the potential at short distance. Indeed, in the range of average interparticle separations explored in this work (namely the 3.5-3.7 \AA \ range), the differences between the interactions are minimal. Also shown in Fig. \ref{fig:pot} is the Silvera-Goldman potential \cite{SG}, which is arguably the most widely adopted in theoretical studies of the condensed phase of H$_2$, and has proven to afford a quantitatively accurate \cite{op} description of structure and energetics of the crystal. As one can see, it has a significantly smaller diameter and is considerably ``softer'' than the potential of Ref. \onlinecite{Szal}; the reason is that it incorporates, in an effective way, non-additive contributions (chiefly triplets), whose overall effect is to soften the repulsive core of the pair-wise part computed {\em ab initio}.
\\ \indent
It therefore seems reasonable to utilize the Silvera-Goldman potential \cite{SG} to carry out our study, as the use of a quantitatively more accurate potential is not likely to affect the conclusions of our study in a significant way. Furthermore, because  the majority of the theoretical studies of condensed H$_2$ have been carried out using the Silvera-Goldman potential, its use in this study allows us to assess the effect of mass alone.
\\ \indent

\section*{Methodology}
The low temperature phase diagram of the thermodynamic system described by Eq.  (\ref{u}) as a function of  density and temperature has been studied in this work by means of  first principles numerical simulations, based on the continuous-space Worm Algorithm \cite{worm1,worm2}.  Since this technique is by now fairly well-established, and extensively described in the literature, we shall not review it here; we used a variant of the algorithm in which the number of particles $N$ is fixed \cite{mezz1,mezz2}.
Details of the simulation are  standard; we made use of the fourth-order approximation for the short imaginary time ($\tau$)  propagator (see, for instance, Ref. \onlinecite{jltp2}), and all of the results presented here are extrapolated to the $\tau\to 0$ limit. We generally found numerical estimates for structural and energetic properties of interest here, obtained with a value of the time step $\tau\sim 3.0\times 10^{-3}$ K$^{-1}$ to be indistinguishable from the extrapolated ones, within the statistical uncertainties of the calculation. 
We carried out simulations of systems typically comprising a number $N$ of particles. In the cluster calculations, $N$ can be chosen arbitrarily. 
For the bulk, the precise value of $N$ 
depends on the type of crystalline structure assumed; typically, $N$ was set to 128 for simulations of body-centered cubic (bcc) and face-centered cubic (fcc) structures, 216 for hexagonal close-packed (hcp). However, we also performed a few targeted simulations with twice as many particles, in order to gauge the quantitative importance of finite size effects. 
\\ \indent
Physical quantities of interest for the bulk calculations include the energy per particle and pressure as a function of density and temperature, i.e., the thermodynamic equation of state in the low temperature limit. We estimated the contribution to the energy and the pressure arising from pairs of particles at distances greater than the largest distance allowed by the size of the simulation cell (i.e., $L/2$), by approximating the pair correlation function $g(r)$ with 1, for $r > L/2$. We have also computed the pair correlation function and the related static structure factor, in order to assess the presence of crystalline order, which can also be  detected through visual inspection of the imaginary-time paths. 
\\ \indent
 We probed for possible superfluid order through the direct calculation of the superfluid fraction using 
 the well-established winding number estimator \cite{pollock}. In order to assess the propensity of the system to develop a superfluid response, and its proximity to a superfluid transition, we also rely on a more indirect criterion, namely we monitor the frequency of cycles of permutations of identical particles involving a significant fraction of the particles in the system. 
While there is no quantitative connection between permutation cycles and the superfluid fraction \cite{mezzacapo08}, a global superfluid phase requires exchanges of macroscopic numbers of particles (see, for instance, Ref. \onlinecite{feynman}).

\section {Results}
\subsection{Bulk}
We simulated crystalline phases of HMu assuming different structures, namely bcc, fcc and hcp. At low temperature, all of these crystals remain stable in the simulation. The energy difference between different crystal structures is typically small, of the order of the statistical errors of our calculation, i.e., few tenths of a K. This is similar to what is observed in H$_2$ \cite{op}). Consequently, we did not attempt to establish what the actual equilibrium structure is, as this is not central to our investigation.
\\
\indent
As a general remark, we note that the estimates of the physical quantities in which we are interested remain unchanged below a temperature $T=2$ K. Thus, results for temperatures lower than 2 K can be considered ground state estimates. 
\begin{figure}[ht]
\centering
\includegraphics[width=\linewidth]{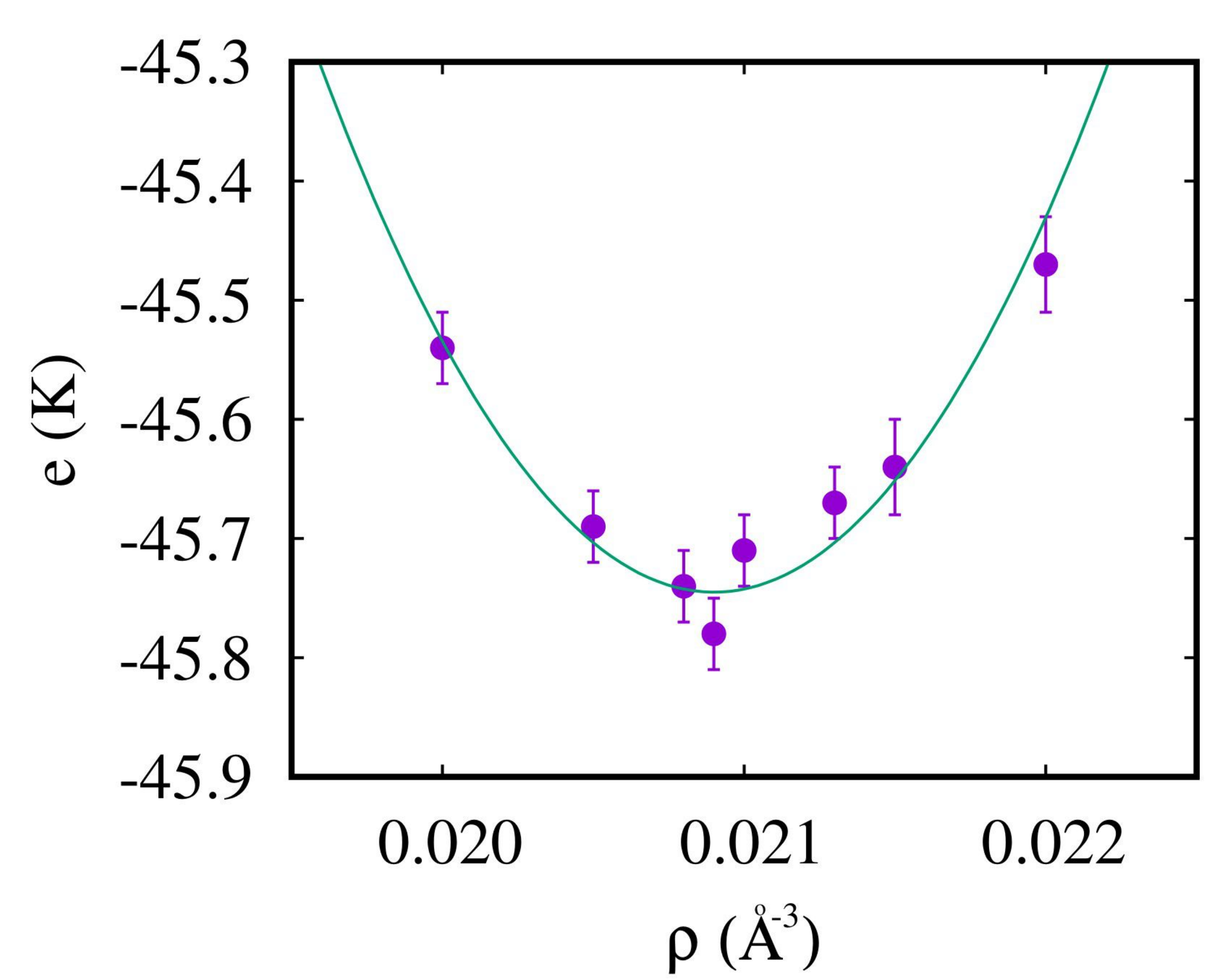}
\caption{Energy per particle $e$ in (K) as a function of density $\rho$ in \AA$^{-3}$, computed at temperature $T=1$ K. Solid line is  a  quadratic fit to the data. These energies are computed assuming a bcc solid structure.}
\label{fig:energy}
\end{figure}
\begin{figure}[ht]
\centering
\includegraphics[width=\linewidth]{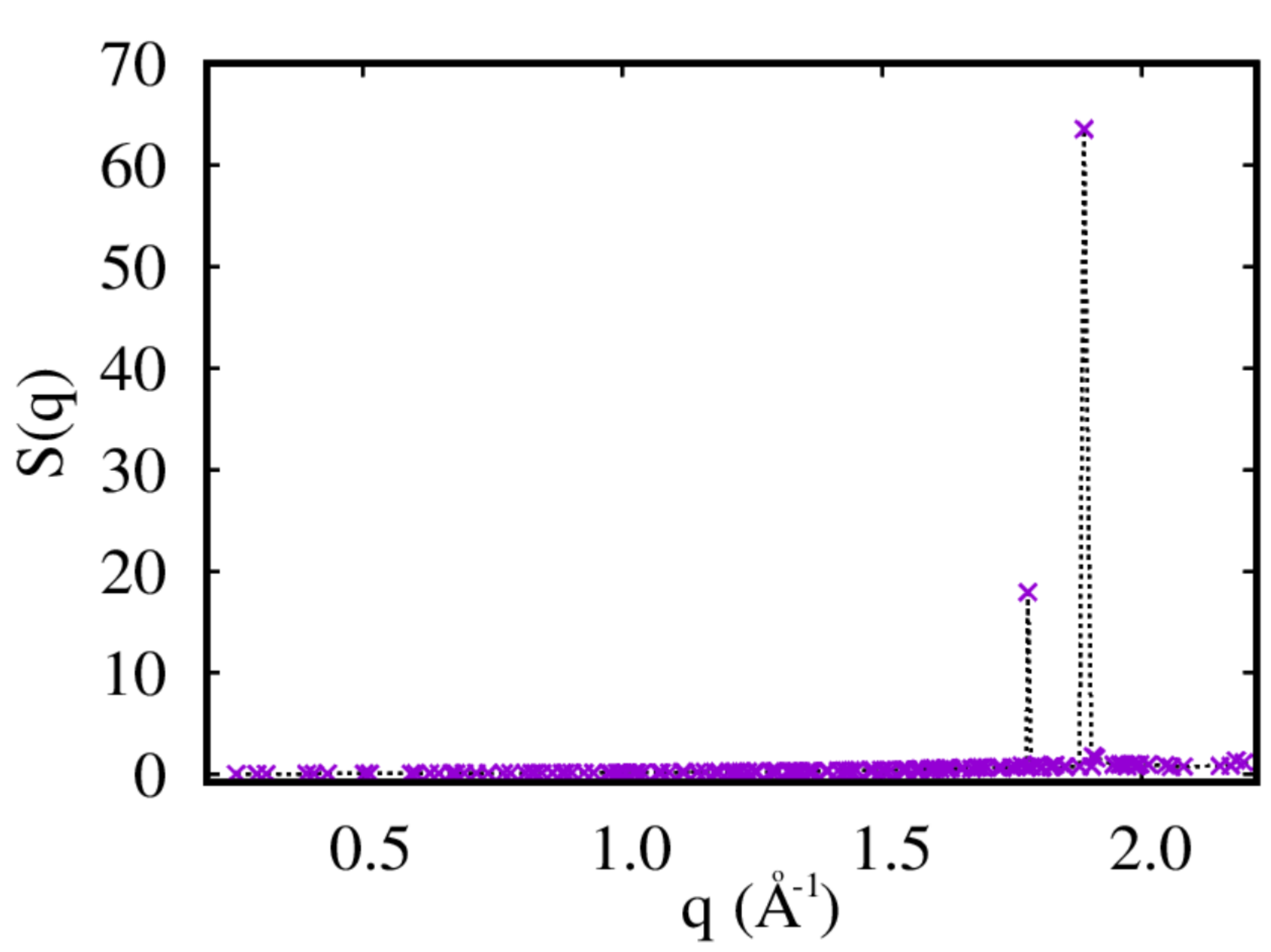}
\caption{Static structure factor computed at temperature $T=1$ K, for a hcp crystal of HMu of density $\rho=0.0209$ \AA$^{-3}$. The simulated system comprises $N=216$ molecules.}
\label{fig:sq}
\end{figure}

We begin by discussing the equation of state of the system in the $T\to 0$ limit.
Fig. \ref{fig:energy} shows the energy per HMu molecule computed as a function of density for a temperature $T=1$ K. The solid line is a quadratic fit to the data, whose fitting parameters yield the equilibrium density $\rho_0=0.02090(5)$ \AA$^{-3}$, corresponding to an average intermolecular distance 3.63 \AA, as well as the ground state energy $e_0=-45.75(2)$ K. The ground state energy is almost exactly \cite{op} one half of that of H$_2$, with a kinetic energy of 70.4(1) K, virtually identical to that of H$_2$ at equilibrium, i.e., at a 30\% higher density. 
\\ \indent
The equilibrium density $\rho_0$ is lower than that of superfluid $^4$He, which is 0.02183 \AA$^{-3}$. 
The system is in the crystalline phase, however, 
as we can ascertain through the calculation of the static structure factor $S(q)$, shown in Fig. \ref{fig:sq}. The result shown in the figure pertains to a case in which particles are initially arranged into a hcp crystal. The sharp peaks occurring at values of $q$ corresponding to wave vectors in the reciprocal lattice signals long-range crystalline order.

\begin{figure}[ht]
\centering
\includegraphics[width=\linewidth]{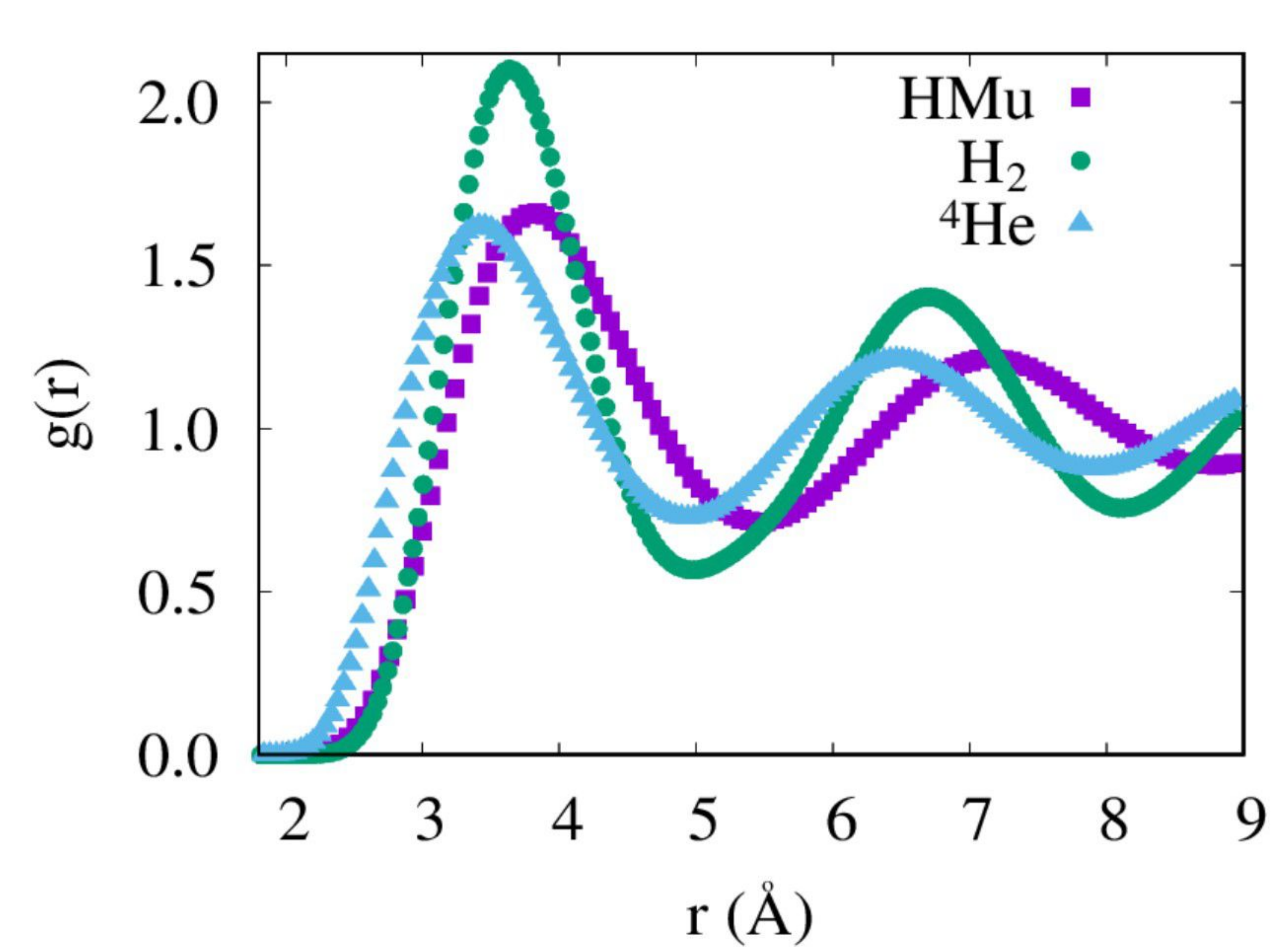}
\caption{Pair correlation function $g(r)$ for HMu at density $\rho_0=0.0209$ \AA$^{-3}$ and $T=1$ K (squares). Also shown for comparison are the corresponding correlation functions for para-H$_2$ at $\rho=0.0261$ \AA$^{-3}$ (circles), and solid $^4$He at $\rho=0.0287$ \AA$^{-3}$ (triangles), at the same temperature. In all three cases the crystalline structure is hcp. Statistical errors are smaller than symbol sizes.}
\label{fig:gr}
\end{figure}
\begin{figure}[ht]
\centering
\includegraphics[width=\linewidth]{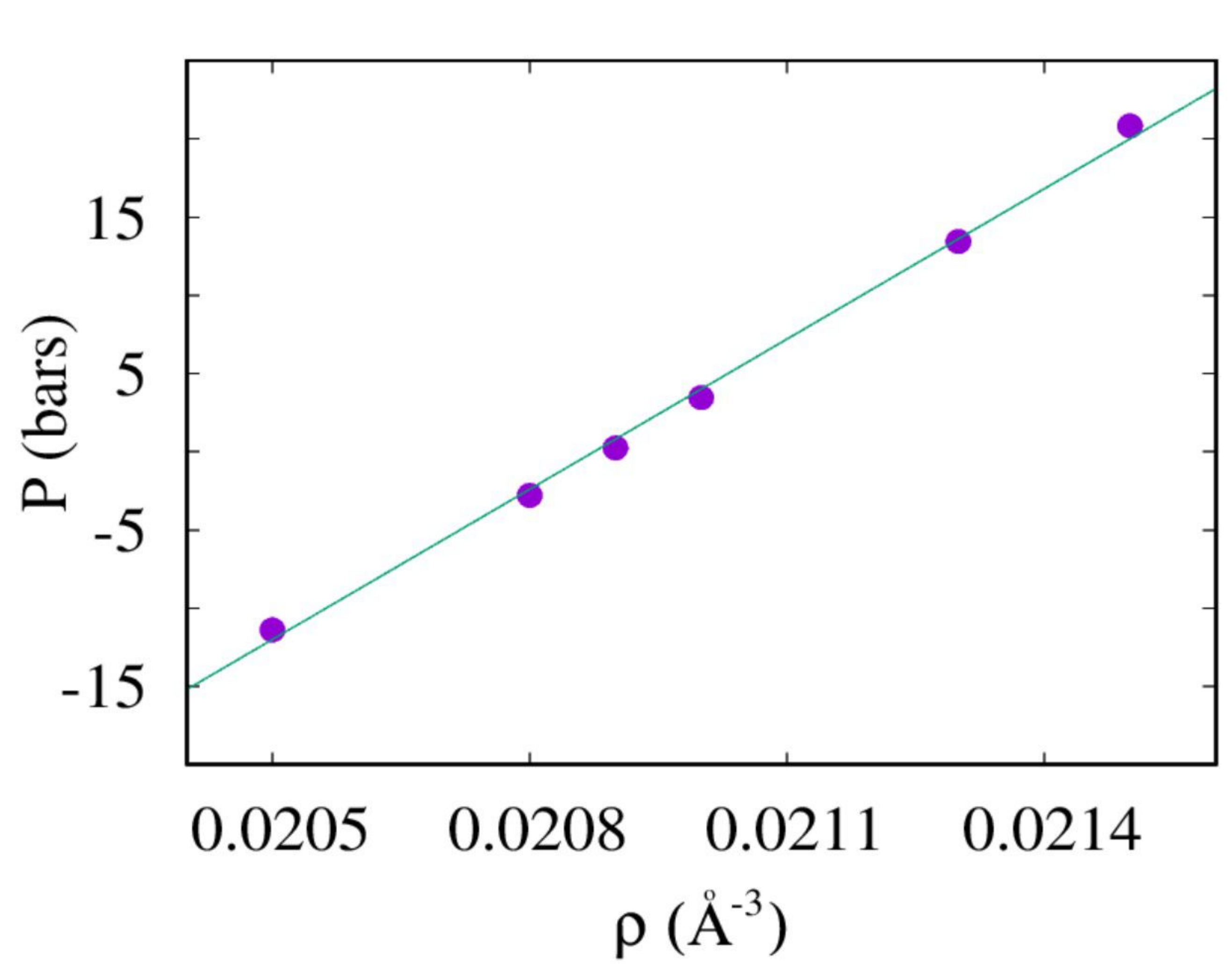}
\caption{Pressure of the HMu bcc crystal as a function of density at 
$T=1$ K (circles). The line is a linear fit to the data. Statistical errors are smaller than symbol sizes.}
\label{fig:press}
\end{figure}

It is interesting to compare the structure of a HMu crystal to that of two other reference quantum solids, namely H$_2$ (more precisely, para-H$_2$) at its equilibrium density, namely $\rho=0.0261$ \AA$^{-3}$, and solid $^4$He near melting, at density $\rho=0.0287$ \AA$^{-3}$. The pair correlation functions shown in Fig. \ref {fig:gr} were computed for hcp crystals at temperature $T=1$ K (the results for H$_2$ and $^4$He were obtained in Ref. \onlinecite{marisa}). While the fact that the peaks appear at different distances reflects the difference in density, $^4$He being the most 
and HMu the least dense crystals, the most noticeable feature is that the height of the peaks is much more pronounced in H$_2$ than in HMu and $^4$He, whose peaks height and widths are 
similar. 
\\ \indent
Fig. \ref{fig:press} shows the pressure of a HMu crystal computed at $T=1$ K, as a function of density. In the relatively narrow density range considered, a linear fit is satisfactory, and allows us to compute the speed of sound $v = (m\rho\kappa)^{-1/2}$, where $\kappa = \rho^{-1}(\partial\rho/\partial P)$ is the compressibility. We 
obtained the speed of sound $1300\pm100$ m/s at the equilibrium density. 
\\ \indent
Next, we discuss the possible superfluid properties of the crystal, as well as of the fluid into which the crystal melts upon raising the temperature. There is no evidence that the crystalline phase of HMu  may display a finite superfluid response in the $T\to 0$ limit. Indeed, the observation is that the behavior of this crystal is virtually identical to that of solid H$_2$, as far as superfluidity is concerned. The main observation is that exchanges of identical particles, which underlie superfluidity, are strongly suppressed in HMu, much like they are in H$_2$, mainly due to the relatively large diameter of the repulsive core of the pairwise interaction. Indeed, exchanges are essentially non-existent in solid HMu at the lowest temperature considered here ($T=1$ K), much like in solid H$_2$; the reduction of particle mass by a factor two does not quantitatively alter the physics of the system in this regard. 
\\ \indent
As temperature is raised, we estimate the melting temperature to be around $T\sim 9$ K. We arrive to that conclusion by computing the pressure as a function of temperature for different densities (at and below $\rho_0$), and by verifying that the system retains solid order even at negative pressure, all the way up to $T=8$ K. No evidence is seen of any metastable, under-pressurized fluid phase; on lowering the density, the system eventually breaks down in solid clusters, just like in H$_2$ \cite{2d}. Above 8 K, the pressure of the solid phase jumps and stable fluid phases appear at lower densities, signaling the occurrence of melting. These fluid phases do not display any superfluid properties; exchanges of two or three particles occur with a frequency of approximately 0.1\%. It is known that {quantum mechanical effects} in H$_2$ contribute about one half of the Lindemann ratio at melting \cite{marisa}; we did not perform the same calculation for this system, but in this system quantum mechanics should be more important, on account of the lighter mass. However, qualitatively melting appears to occur very similarly as in H$_2$.
Also, our results suggest that thermal expansion, which is negligible \cite{cabrillo} in solid H$_2$, is  likely very small in this crystal.
\subsection{Clusters}
It is also interesting to study the physics of nanoscale size clusters of HMu, and compare their behavior to that of parahydrogen clusters, for which, as mentioned above, superfluid behavior is predicted at temperature of the order of 1 K, if their size is approximately 30 molecules or less. Crystalline behavior emerges rather rapidly for parahydrogen clusters of greater size \cite{cqp}, with ``supersolid'' behavior occurring for specific clusters \cite{supercl}. This calculation is carried out with the same methodology adopted for the bulk phase of the system, the only difference being that the simulation cell is now taken large enough that a single cluster forms. In this respect, the behavior of HMu clusters is closer to that of parahydrogen than $^4$He clusters,  in that no external potential is required to keep them together (as in the case of $^4$He), at least at sufficiently low temperature ($\lesssim 4$ K).
\\ \indent
However, the reduced molecular mass makes the physics of HMu clusters both quantitatively and qualitatively different from that of parahydrogen clusters. The first observation is that the superfluid response is greatly enhanced; specifically, it is found that clusters with as many as $N=200$ molecules are close to 100\% superfluid at $T=1$ K, and remain at least 50\% superfluid up to $T\lesssim4$ K, at which point they begin to evaporate, i.e., they do not appear to stay together as normal clusters. 
This is in stark contrast with parahydrogen clusters, where exchanges are rare in clusters of more than 30 molecules at this temperature, and they would involve at most $\sim 10$ particles. Consistently, the superfluid response is insignificant in parahydrogen clusters, and they stay together almost exclusively due to the potential energy of interaction.
Our results highlight the importance of  the energetic contribution of quantum-mechanical exchanges in the stabilization of HMu clusters at low temperature. Indeed, at $T=1$ K we observe cycles of exchanges involving {\em all} of the particles in HMu clusters comprising as many as 200 molecules. 
\begin{figure}[ht]
\centering
\includegraphics[width=\linewidth]{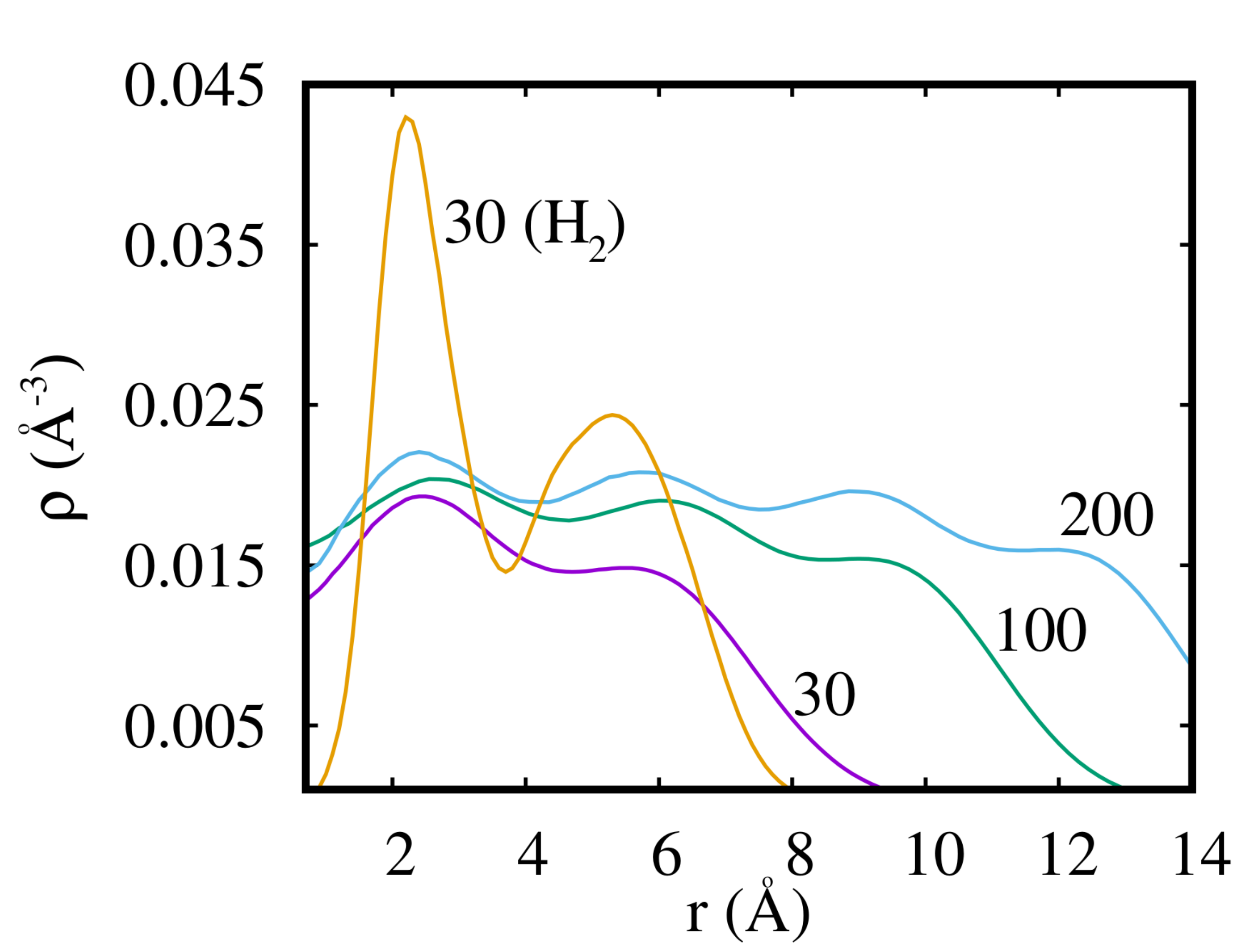}
\caption{Radial density profiles (computed with respect to the center of mass) for HMu clusters comprising $N=30$, 100 and 200 molecules, at temperature $T=1$ K. At this temperature, all of these clusters are essentially 100\% superfluid. Also shown for comparison is the same quantity for a cluster of $N=30$ parahydrogen molecules, at the same temperature; this cluster has no significant superfluid response.}
\label{profiles}
\end{figure}
\\ \indent
Fig. \ref{profiles} shows  density profiles computed with respect to the
center of mass
of the system for three different clusters of HMu, comprising $N=30$, 100 and 200 molecules. As mentioned above, all these clusters are essentially fully superfluid at this temperature. These  density profiles are qualitatively similar, nearly featureless and reminiscent of those computed for $^4$He droplets \cite{sindzingre89}. The comparison with the density profile computed for a cluster of 30 parahydrogen molecules, also shown in Fig. \ref{profiles}, illustrates how the latter is much more compact and displays pronounced oscillations, which are indicative of a well-defined, solid-like shell structure. \\ \indent 
As the number of molecules increases, clusters of HMu ought to evolve into solid-like objects, their structure approaching that of the bulk. The calculation of the radial superfluid density \cite{kwonls,mezzacapo08} suggests that crystallization begins to occur at the center of the cluster; for example, at $T=1$ K the largest HMu cluster studied here ($N=200$) displays a suppressed superfluid response inside a central core of approximately 5 \AA\ radius, in which crystalline order slowly begins to emerge, while the rest of the cluster is essentially entirely superfluid. We expect the non-superfluid core to grow with the size of the cluster. In other words, large clusters consist of a rigid, insulating core and a superfluid surface layer, with a rather clear demarcation between the two. This is qualitatively different from the behavior observed in parahydrogen clusters, in which crystallization occurs for much smaller sizes, and ``supersolid'' clusters simultaneously displaying  solid-like and superfluid properties can be identified \cite{supercl}.  
\section{Conclusions}
We have investigated the low temperature phase diagram of a bulk assembly of muonium hydride molecules, by means of first principle quantum simulations. Our model assumes a pairwise, central interaction among HMu molecules which is identical to that of H$_2$ molecules.
By a mapping of the H$_2$ inter-atomic potentials derived from ab initio 
calculations, we showed that this model provides a reasonable description 
of the HMu-HMu interation.
It is certainly possible to carry out a more accurate determination of the pair potential, but 
the main effect of the lower $\mu^+$ mass should be that of increasing the diameter of the repulsive core of the interaction, in turn suppressing exchanges even more. As  illustrated in Ref. \onlinecite{role}, exchanges of identical particles play a crucial role in destabilizing the  classical picture in a many-body system; when exchanges are suppressed, quantum zero point motion can only alter such a picture quantitatively, not qualitatively. Alternatively, this can be understood by the fact that $\Lambda$ is decreased
as the hard core radius is increased, making the system more classical. Obviously, regarding the interaction as spherically symmetric is also an approximation, but one that affords quantitatively accurate results for parahydrogen \cite{op}, for which the potential energy of interaction among molecules plays a quantitatively more important role in shaping the phase diagram of the condensed system.
\\ \indent
Perhaps the most significant observation of this study is the different physics of bulk and nanoscale size clusters of HMu. Bulk HMu is very similar to H$_2$; despite its very low density (lower than that of superfluid $^4$He), the equilibrium crystalline phase is stable below a temperature of about 9 K, much closer to the melting temperature of H$_2$ (13.8 K) than the mass difference may have led one to expect. 
No evidence is observed of any superfluid phase, neither liquid nor crystalline, underscoring once again how, in order for a supersolid phase to be possible in continuous space, { it requires some physical mechanism to cause a ``softening" of the} repulsive part of the pair potential at short distances \cite{su}, even if only along one direction, like in the case of a dipolar interaction \cite{patterned}. On the other hand, clusters of HMu including up to a few hundred molecules display superfluid behavior similar to that of $^4$He clusters. This suggests that, as the value of the quantumness parameter $\Lambda$ approaches $\Lambda_c$ from below, one may observe nanoscale superfluidity in clusters of rather large size.
\\ \indent
We conclude by discussing the possible experimental realization of the system described in this work. The replacement of elementary constituents of matter, typically electrons, with other subatomic particles of the same charge, such as muons \cite{Egan}, has been discussed for a long time, and some experimental success has been reported. Even a bold scenario consisting of  replacing {\em all} electrons \cite{wheeler} in atoms with muons (the so-called ``muonic matter'') has been considered; recently a long-lived ``pionic helium'' has been created \cite{pionic}.  Thus, 
it also seems plausible to replace  a proton in a H$_2$ molecule with an antimuon; indeed, muonium chemistry has been an active area of research for several decades \cite{muonium}. 
In order for ``muonium condensed matter" to be feasible, 
a main challenge to overcome is the very short lifetime of the $\mu^+$, of the order of a $\mu$s. 

\section*{Acknowledgements}
This work was supported by the Natural Sciences and Engineering Research Council of Canada,
a Simons Investigator grant (DTS) and the Simons Collaboration on Ultra-Quantum Matter, which is a grant from the Simons Foundation (651440, DTS).
Computing support of Compute Canada
and of the Flatiron Institute 
are
gratefully acknowledged. 
The Flatiron Institute is a division of the Simons Foundation.

\bibliography{sample}

\begin{thebibliography}{38}%
\makeatletter
\providecommand \@ifxundefined [1]{%
 \@ifx{#1\undefined}
}%
\providecommand \@ifnum [1]{%
 \ifnum #1\expandafter \@firstoftwo
 \else \expandafter \@secondoftwo
 \fi
}%
\providecommand \@ifx [1]{%
 \ifx #1\expandafter \@firstoftwo
 \else \expandafter \@secondoftwo
 \fi
}%
\providecommand \natexlab [1]{#1}%
\providecommand \enquote  [1]{``#1''}%
\providecommand \bibnamefont  [1]{#1}%
\providecommand \bibfnamefont [1]{#1}%
\providecommand \citenamefont [1]{#1}%
\providecommand \href@noop [0]{\@secondoftwo}%
\providecommand \href [0]{\begingroup \@sanitize@url \@href}%
\providecommand \@href[1]{\@@startlink{#1}\@@href}%
\providecommand \@@href[1]{\endgroup#1\@@endlink}%
\providecommand \@sanitize@url [0]{\catcode `\\12\catcode `\$12\catcode
  `\&12\catcode `\#12\catcode `\^12\catcode `\_12\catcode `\%12\relax}%
\providecommand \@@startlink[1]{}%
\providecommand \@@endlink[0]{}%
\providecommand \url  [0]{\begingroup\@sanitize@url \@url }%
\providecommand \@url [1]{\endgroup\@href {#1}{\urlprefix }}%
\providecommand \urlprefix  [0]{URL }%
\providecommand \Eprint [0]{\href }%
\providecommand \doibase [0]{http://dx.doi.org/}%
\providecommand \selectlanguage [0]{\@gobble}%
\providecommand \bibinfo  [0]{\@secondoftwo}%
\providecommand \bibfield  [0]{\@secondoftwo}%
\providecommand \translation [1]{[#1]}%
\providecommand \BibitemOpen [0]{}%
\providecommand \bibitemStop [0]{}%
\providecommand \bibitemNoStop [0]{.\EOS\space}%
\providecommand \EOS [0]{\spacefactor3000\relax}%
\providecommand \BibitemShut  [1]{\csname bibitem#1\endcsname}%
\let\auto@bib@innerbib\@empty
\bibitem [{\citenamefont {Kora}\ \emph {et~al.}(2020)\citenamefont {Kora},
  \citenamefont {Boninsegni}, \citenamefont {Son},\ and\ \citenamefont
  {Zhang}}]{pnas}%
  \BibitemOpen
  \bibfield  {author} {\bibinfo {author} {\bibfnamefont {Y.}~\bibnamefont
  {Kora}}, \bibinfo {author} {\bibfnamefont {M.}~\bibnamefont {Boninsegni}},
  \bibinfo {author} {\bibfnamefont {D.~T.}\ \bibnamefont {Son}}, \ and\
  \bibinfo {author} {\bibfnamefont {S.}~\bibnamefont {Zhang}},\ }\bibfield
  {title} {\enquote {\bibinfo {title} {Tuning the quantunmness of simple {Bose}
  systems: A universal phase diagram},}\ }\href {\doibase
  https://doi.org/10.1073/pnas.2017646117} {\bibfield  {journal} {\bibinfo
  {journal} {Proc. Natl. Acad. Sci.}\ }\textbf {\bibinfo {volume} {117}},\
  \bibinfo {pages} {27231--27237} (\bibinfo {year} {2020})}\BibitemShut
  {NoStop}%
\bibitem [{\citenamefont {Nosanow}\ \emph {et~al.}(1975)\citenamefont
  {Nosanow}, \citenamefont {Parish},\ and\ \citenamefont {Pinski}}]{nosanow}%
  \BibitemOpen
  \bibfield  {author} {\bibinfo {author} {\bibfnamefont {L.~H.}\ \bibnamefont
  {Nosanow}}, \bibinfo {author} {\bibfnamefont {L.~J.}\ \bibnamefont {Parish}},
  \ and\ \bibinfo {author} {\bibfnamefont {F.~J.}\ \bibnamefont {Pinski}},\
  }\bibfield  {title} {\enquote {\bibinfo {title} {Zero-temperature properties
  of matter and the quantum theorem of corresponding states: The
  liquid-to-crystal phase transition for {Fermi and Bose} systems},}\ }\href
  {\doibase http://dx.doi.org/10.1103/PhysRevB.11.191} {\bibfield  {journal}
  {\bibinfo  {journal} {Phys. Rev. B}\ }\textbf {\bibinfo {volume} {11}},\
  \bibinfo {pages} {191--204} (\bibinfo {year} {1975})}\BibitemShut {NoStop}%
\bibitem [{\citenamefont {Ginzburg}\ and\ \citenamefont
  {Sobyanin}(1972)}]{ginzburg}%
  \BibitemOpen
  \bibfield  {author} {\bibinfo {author} {\bibfnamefont {V.~L.}\ \bibnamefont
  {Ginzburg}}\ and\ \bibinfo {author} {\bibfnamefont {A.~A.}\ \bibnamefont
  {Sobyanin}},\ }\bibfield  {title} {\enquote {\bibinfo {title} {Can liquid
  molecular-hydrogen be superfluid?}}\ }\href@noop {} {\bibfield  {journal}
  {\bibinfo  {journal} {JETP Letters-USSR}\ }\textbf {\bibinfo {volume} {15}},\
  \bibinfo {pages} {242} (\bibinfo {year} {1972})}\BibitemShut {NoStop}%
\bibitem [{\citenamefont {Boninsegni}(2018)}]{boninsegni18}%
  \BibitemOpen
  \bibfield  {author} {\bibinfo {author} {\bibfnamefont {M.}~\bibnamefont
  {Boninsegni}},\ }\bibfield  {title} {\enquote {\bibinfo {title} {Search for
  superfluidity in supercooled liquid parahydrogen},}\ }\href {\doibase
  http://dx.doi.org/10.1103/PhysRevB.97.054517} {\bibfield  {journal} {\bibinfo
   {journal} {{Phys. Rev. B}}\ }\textbf {\bibinfo {volume} {97}},\ \bibinfo
  {pages} {054517} (\bibinfo {year} {2018})}\BibitemShut {NoStop}%
\bibitem [{\citenamefont {Boninsegni}(2004)}]{2d}%
  \BibitemOpen
  \bibfield  {author} {\bibinfo {author} {\bibfnamefont {M.}~\bibnamefont
  {Boninsegni}},\ }\bibfield  {title} {\enquote {\bibinfo {title}
  {{Low-temperature phase diagram of condensed para-hydrogen in two
  dimensions}},}\ }\href {\doibase 10.1103/PhysRevB.70.193411} {\bibfield
  {journal} {\bibinfo  {journal} {Phys. Rev. B}\ }\textbf {\bibinfo {volume}
  {70}},\ \bibinfo {pages} {193411} (\bibinfo {year} {2004})}\BibitemShut
  {NoStop}%
\bibitem [{\citenamefont {Boninsegni}(2013)}]{1d}%
  \BibitemOpen
  \bibfield  {author} {\bibinfo {author} {\bibfnamefont {M.}~\bibnamefont
  {Boninsegni}},\ }\bibfield  {title} {\enquote {\bibinfo {title} {{Ground
  State Phase Diagram of Parahydrogen in One Dimension}},}\ }\href {\doibase
  10.1103/PhysRevLett.111.235303} {\bibfield  {journal} {\bibinfo  {journal}
  {Phys. Rev. Lett.}\ }\textbf {\bibinfo {volume} {111}},\ \bibinfo {pages}
  {235303} (\bibinfo {year} {2013})}\BibitemShut {NoStop}%
\bibitem [{\citenamefont {Sindzingre}\ \emph {et~al.}(1991)\citenamefont
  {Sindzingre}, \citenamefont {Ceperley},\ and\ \citenamefont
  {Klein}}]{sindzingre}%
  \BibitemOpen
  \bibfield  {author} {\bibinfo {author} {\bibfnamefont {P.}~\bibnamefont
  {Sindzingre}}, \bibinfo {author} {\bibfnamefont {D.~M.}\ \bibnamefont
  {Ceperley}}, \ and\ \bibinfo {author} {\bibfnamefont {M.~L.}\ \bibnamefont
  {Klein}},\ }\bibfield  {title} {\enquote {\bibinfo {title} {{Superfluidity in
  Clusters of p-H$_2$ molecules}},}\ }\href {\doibase
  10.1103/PhysRevLett.67.1871} {\bibfield  {journal} {\bibinfo  {journal}
  {Phys. Rev. Lett.}\ }\textbf {\bibinfo {volume} {67}},\ \bibinfo {pages}
  {1871--1874} (\bibinfo {year} {1991})}\BibitemShut {NoStop}%
\bibitem [{\citenamefont {Mezzacapo}\ and\ \citenamefont
  {Boninsegni}(2006)}]{mezz1}%
  \BibitemOpen
  \bibfield  {author} {\bibinfo {author} {\bibfnamefont {F.}~\bibnamefont
  {Mezzacapo}}\ and\ \bibinfo {author} {\bibfnamefont {M.}~\bibnamefont
  {Boninsegni}},\ }\bibfield  {title} {\enquote {\bibinfo {title}
  {{Superfluidity and Quantum Melting of p-H$_2$ Clusters}},}\ }\href {\doibase
  10.1103/physrevlett.97.045301} {\bibfield  {journal} {\bibinfo  {journal}
  {Phys. Rev. Lett.}\ }\textbf {\bibinfo {volume} {97}},\ \bibinfo {pages}
  {045301} (\bibinfo {year} {2006})}\BibitemShut {NoStop}%
\bibitem [{\citenamefont {Mezzacapo}\ and\ \citenamefont
  {Boninsegni}(2007)}]{mezz2}%
  \BibitemOpen
  \bibfield  {author} {\bibinfo {author} {\bibfnamefont {F.}~\bibnamefont
  {Mezzacapo}}\ and\ \bibinfo {author} {\bibfnamefont {M.}~\bibnamefont
  {Boninsegni}},\ }\bibfield  {title} {\enquote {\bibinfo {title} {Structure,
  superfluidity, and quantum melting of hydrogen clusters},}\ }\href {\doibase
  10.1103/physreva.75.033201} {\bibfield  {journal} {\bibinfo  {journal} {Phys.
  Rev. A}\ }\textbf {\bibinfo {volume} {75}},\ \bibinfo {pages} {033201}
  (\bibinfo {year} {2007})}\BibitemShut {NoStop}%
\bibitem [{\citenamefont {Boninsegni}(2020)}]{2020}%
  \BibitemOpen
  \bibfield  {author} {\bibinfo {author} {\bibfnamefont {M.}~\bibnamefont
  {Boninsegni}},\ }\bibfield  {title} {\enquote {\bibinfo {title} {{Superfluid
  Response of Parahydrogen Clusters in Superfluid $^4$He}},}\ }\href {\doibase
  10.1007/s10909-020-02493-4} {\bibfield  {journal} {\bibinfo  {journal} {J.
  Low Temp. Phys.}\ }\textbf {\bibinfo {volume} {201}},\ \bibinfo {pages}
  {193--199} (\bibinfo {year} {2020})}\BibitemShut {NoStop}%
\bibitem [{\citenamefont {Grebenev}\ \emph {et~al.}(2000)\citenamefont
  {Grebenev}, \citenamefont {Sartakov}, \citenamefont {Toennies},\ and\
  \citenamefont {Vilesov}}]{grebenev}%
  \BibitemOpen
  \bibfield  {author} {\bibinfo {author} {\bibfnamefont {S.}~\bibnamefont
  {Grebenev}}, \bibinfo {author} {\bibfnamefont {B.}~\bibnamefont {Sartakov}},
  \bibinfo {author} {\bibfnamefont {J.~P.}\ \bibnamefont {Toennies}}, \ and\
  \bibinfo {author} {\bibfnamefont {A.~F.}\ \bibnamefont {Vilesov}},\
  }\bibfield  {title} {\enquote {\bibinfo {title} {{Evidence for Superﬂuidity
  in Para-Hydrogen Clusters Inside Helium-4 Droplets at 0.15 Kelvin}},}\ }\href
  {\doibase 10.1007/s10909-020-02493-4} {\bibfield  {journal} {\bibinfo
  {journal} {Science}\ }\textbf {\bibinfo {volume} {289}},\ \bibinfo {pages}
  {1532--1535} (\bibinfo {year} {2000})}\BibitemShut {NoStop}%
\bibitem [{\citenamefont {Boninsegni}\ and\ \citenamefont
  {Prokof'ev}(2012)}]{supersolid}%
  \BibitemOpen
  \bibfield  {author} {\bibinfo {author} {\bibfnamefont {M.}~\bibnamefont
  {Boninsegni}}\ and\ \bibinfo {author} {\bibfnamefont {N.~V.}\ \bibnamefont
  {Prokof'ev}},\ }\bibfield  {title} {\enquote {\bibinfo {title} {Supersolids:
  What and where are they?}}\ }\href {\doibase
  https://doi.org/10.1103/RevModPhys.84.759} {\bibfield  {journal} {\bibinfo
  {journal} {{Rev. Mod. Phys.}}\ }\textbf {\bibinfo {volume} {84}},\ \bibinfo
  {pages} {759--776} (\bibinfo {year} {2012})}\BibitemShut {NoStop}%
\bibitem [{Note1()}]{Note1}%
  \BibitemOpen
  \bibinfo {note} {Specifically, the average distance between the proton and
  the muon in a HMu molecule is estimated at 0.8 \r A, whereas that between the
  two protons in a H$_2$ molecule is 0.74 \r A. See Refs, \cite
  {suff,zhou}}\BibitemShut {NoStop}%
\bibitem [{\citenamefont {Feynman}(1953)}]{feynman}%
  \BibitemOpen
  \bibfield  {author} {\bibinfo {author} {\bibfnamefont {R.~P.}\ \bibnamefont
  {Feynman}},\ }\bibfield  {title} {\enquote {\bibinfo {title} {{Atomic theory
  of the $\lambda$ transition in helium}},}\ }\href {\doibase
  https://doi.org/10.1103/PhysRev.91.1291} {\bibfield  {journal} {\bibinfo
  {journal} {Phys. Rev.}\ }\textbf {\bibinfo {volume} {91}},\ \bibinfo {pages}
  {1291--1301} (\bibinfo {year} {1953})}\BibitemShut {NoStop}%
\bibitem [{\citenamefont {Boninsegni}\ \emph {et~al.}(2012)\citenamefont
  {Boninsegni}, \citenamefont {Pollet}, \citenamefont {Prokof'ev},\ and\
  \citenamefont {Svistunov}}]{role}%
  \BibitemOpen
  \bibfield  {author} {\bibinfo {author} {\bibfnamefont {M.}~\bibnamefont
  {Boninsegni}}, \bibinfo {author} {\bibfnamefont {L.}~\bibnamefont {Pollet}},
  \bibinfo {author} {\bibfnamefont {N.}~\bibnamefont {Prokof'ev}}, \ and\
  \bibinfo {author} {\bibfnamefont {B.}~\bibnamefont {Svistunov}},\ }\bibfield
  {title} {\enquote {\bibinfo {title} {{Role of {Bose} statistics in
  crystallization and quantum jamming}},}\ }\href {\doibase
  https://doi.org/10.1103/PhysRevLett.109.025302} {\bibfield  {journal}
  {\bibinfo  {journal} {Phys. Rev. Lett.}\ }\textbf {\bibinfo {volume} {109}},\
  \bibinfo {pages} {025302} (\bibinfo {year} {2012})}\BibitemShut {NoStop}%
\bibitem [{\citenamefont {Silvera}\ and\ \citenamefont {Goldman}(1978)}]{SG}%
  \BibitemOpen
  \bibfield  {author} {\bibinfo {author} {\bibfnamefont {I.~F.}\ \bibnamefont
  {Silvera}}\ and\ \bibinfo {author} {\bibfnamefont {V.~V.}\ \bibnamefont
  {Goldman}},\ }\bibfield  {title} {\enquote {\bibinfo {title} {{The isotropic
  intermolecular potential for H2 and D2 in the solid and gas phases}},}\
  }\href {\doibase https://doi.org/ https://doi.org/10.1063/1.437103}
  {\bibfield  {journal} {\bibinfo  {journal} {J. Chem. Phys.}\ }\textbf
  {\bibinfo {volume} {69}},\ \bibinfo {pages} {4209--4217} (\bibinfo {year}
  {1978})}\BibitemShut {NoStop}%
\bibitem [{\citenamefont {Diep}\ and\ \citenamefont {Johnson}(2000)}]{DJ}%
  \BibitemOpen
  \bibfield  {author} {\bibinfo {author} {\bibfnamefont {P.}~\bibnamefont
  {Diep}}\ and\ \bibinfo {author} {\bibfnamefont {J.~K.}\ \bibnamefont
  {Johnson}},\ }\bibfield  {title} {\enquote {\bibinfo {title} {{An accurate
  H$_2$–H$_2$ interaction potential from first principles}},}\ }\href
  {\doibase https://doi.org/ https://doi.org/10.1063/1.481009} {\bibfield
  {journal} {\bibinfo  {journal} {J. Chem. Phys.}\ }\textbf {\bibinfo {volume}
  {112}},\ \bibinfo {pages} {4465--4473} (\bibinfo {year} {2000})}\BibitemShut
  {NoStop}%
\bibitem [{\citenamefont {Patkowski}\ \emph {et~al.}(2008)\citenamefont
  {Patkowski}, \citenamefont {Cencek}, \citenamefont {Jankowski}, \citenamefont
  {Szalewicz}, \citenamefont {Mehl}, \citenamefont {Garberoglio},\ and\
  \citenamefont {Harvey}}]{Szal}%
  \BibitemOpen
  \bibfield  {author} {\bibinfo {author} {\bibfnamefont {K.}~\bibnamefont
  {Patkowski}}, \bibinfo {author} {\bibfnamefont {W.}~\bibnamefont {Cencek}},
  \bibinfo {author} {\bibfnamefont {P.}~\bibnamefont {Jankowski}}, \bibinfo
  {author} {\bibfnamefont {K.}~\bibnamefont {Szalewicz}}, \bibinfo {author}
  {\bibfnamefont {J.~B.}\ \bibnamefont {Mehl}}, \bibinfo {author}
  {\bibfnamefont {G.}~\bibnamefont {Garberoglio}}, \ and\ \bibinfo {author}
  {\bibfnamefont {A.~H.}\ \bibnamefont {Harvey}},\ }\bibfield  {title}
  {\enquote {\bibinfo {title} {{Potential energy surface for interactions
  between two hydrogen molecules}},}\ }\href {\doibase https://doi.org/
  https://doi.org/10.1063/1.481009} {\bibfield  {journal} {\bibinfo  {journal}
  {J. Chem. Phys.}\ }\textbf {\bibinfo {volume} {129}},\ \bibinfo {pages}
  {094304} (\bibinfo {year} {2008})}\BibitemShut {NoStop}%
\bibitem [{\citenamefont {Zhou}\ \emph {et~al.}(2005)\citenamefont {Zhou},
  \citenamefont {Zhu},\ and\ \citenamefont {Yan}}]{zhou}%
  \BibitemOpen
  \bibfield  {author} {\bibinfo {author} {\bibfnamefont {B.-L.}\ \bibnamefont
  {Zhou}}, \bibinfo {author} {\bibfnamefont {J.-M.}\ \bibnamefont {Zhu}}, \
  and\ \bibinfo {author} {\bibfnamefont {Z.-C.}\ \bibnamefont {Yan}},\
  }\bibfield  {title} {\enquote {\bibinfo {title} {Variational calculation of
  muonium hydride},}\ }\href {\doibase
  https://doi.org/10.1088/0953-4075/38/3/014} {\bibfield  {journal} {\bibinfo
  {journal} {{J. Phys. B: At. Mol. Opt. Phys.}}\ }\textbf {\bibinfo {volume}
  {38}},\ \bibinfo {pages} {305--309} (\bibinfo {year} {2005})}\BibitemShut
  {NoStop}%
\bibitem [{\citenamefont {Operetto}\ and\ \citenamefont {Pederiva}(2006)}]{op}%
  \BibitemOpen
  \bibfield  {author} {\bibinfo {author} {\bibfnamefont {F.}~\bibnamefont
  {Operetto}}\ and\ \bibinfo {author} {\bibfnamefont {F.}~\bibnamefont
  {Pederiva}},\ }\bibfield  {title} {\enquote {\bibinfo {title} {Diffusion
  monte carlo study of the equation of state of solid para-h$_2$},}\ }\href
  {\doibase 10.1103/PhysRevB.73.184124} {\bibfield  {journal} {\bibinfo
  {journal} {Phys. Rev. B}\ }\textbf {\bibinfo {volume} {73}},\ \bibinfo
  {pages} {184124} (\bibinfo {year} {2006})}\BibitemShut {NoStop}%
\bibitem [{\citenamefont {Boninsegni}\ \emph
  {et~al.}(2006{\natexlab{a}})\citenamefont {Boninsegni}, \citenamefont
  {Prokof'ev},\ and\ \citenamefont {Svistunov}}]{worm1}%
  \BibitemOpen
  \bibfield  {author} {\bibinfo {author} {\bibfnamefont {M.}~\bibnamefont
  {Boninsegni}}, \bibinfo {author} {\bibfnamefont {N.~V.}\ \bibnamefont
  {Prokof'ev}}, \ and\ \bibinfo {author} {\bibfnamefont {B.~V.}\ \bibnamefont
  {Svistunov}},\ }\bibfield  {title} {\enquote {\bibinfo {title} {{Worm
  Algorithm for Continuous-Space Path Integral Monte Carlo Simulations}},}\
  }\href {\doibase 10.1103/physrevlett.96.070601} {\bibfield  {journal}
  {\bibinfo  {journal} {Phys. Rev. Lett.}\ }\textbf {\bibinfo {volume} {96}},\
  \bibinfo {pages} {070601} (\bibinfo {year} {2006}{\natexlab{a}})}\BibitemShut
  {NoStop}%
\bibitem [{\citenamefont {Boninsegni}\ \emph
  {et~al.}(2006{\natexlab{b}})\citenamefont {Boninsegni}, \citenamefont
  {Prokof'ev},\ and\ \citenamefont {Svistunov}}]{worm2}%
  \BibitemOpen
  \bibfield  {author} {\bibinfo {author} {\bibfnamefont {M.}~\bibnamefont
  {Boninsegni}}, \bibinfo {author} {\bibfnamefont {N.~V.}\ \bibnamefont
  {Prokof'ev}}, \ and\ \bibinfo {author} {\bibfnamefont {B.~V.}\ \bibnamefont
  {Svistunov}},\ }\bibfield  {title} {\enquote {\bibinfo {title} {{Worm
  Algorithm and Diagrammatic Monte Carlo: a New Approach to Continuous-Space
  Path Integral Monte Carlo Simulations}},}\ }\href {\doibase
  10.1103/PhysRevE.74.036701} {\bibfield  {journal} {\bibinfo  {journal} {Phys.
  Rev. E}\ }\textbf {\bibinfo {volume} {74}},\ \bibinfo {pages} {036701}
  (\bibinfo {year} {2006}{\natexlab{b}})}\BibitemShut {NoStop}%
\bibitem [{\citenamefont {Boninsegni}(2005)}]{jltp2}%
  \BibitemOpen
  \bibfield  {author} {\bibinfo {author} {\bibfnamefont {M.}~\bibnamefont
  {Boninsegni}},\ }\bibfield  {title} {\enquote {\bibinfo {title} {{Permutation
  sampling in path integral Monte Carlo}},}\ }\href@noop {} {\bibfield
  {journal} {\bibinfo  {journal} {{J. Low Temp. Phys.}}\ }\textbf {\bibinfo
  {volume} {141}},\ \bibinfo {pages} {27--46} (\bibinfo {year}
  {2005})}\BibitemShut {NoStop}%
\bibitem [{\citenamefont {Pollock}\ and\ \citenamefont
  {Ceperley}(1987)}]{pollock}%
  \BibitemOpen
  \bibfield  {author} {\bibinfo {author} {\bibfnamefont {E.~L.}\ \bibnamefont
  {Pollock}}\ and\ \bibinfo {author} {\bibfnamefont {D.~M.}\ \bibnamefont
  {Ceperley}},\ }\bibfield  {title} {\enquote {\bibinfo {title} {Path-integral
  computation of superfluid densities},}\ }\href {\doibase
  10.1103/PhysRevB.36.8343} {\bibfield  {journal} {\bibinfo  {journal} {Phys.
  Rev. B}\ }\textbf {\bibinfo {volume} {36}},\ \bibinfo {pages} {8343--8352}
  (\bibinfo {year} {1987})}\BibitemShut {NoStop}%
\bibitem [{\citenamefont {Mezzacapo}\ and\ \citenamefont
  {Boninsegni}(2008)}]{mezzacapo08}%
  \BibitemOpen
  \bibfield  {author} {\bibinfo {author} {\bibfnamefont {F.}~\bibnamefont
  {Mezzacapo}}\ and\ \bibinfo {author} {\bibfnamefont {M.}~\bibnamefont
  {Boninsegni}},\ }\bibfield  {title} {\enquote {\bibinfo {title} {Local
  superfluidity of parahydrogen clusters},}\ }\href {\doibase
  10.1103/PhysRevLett.100.145301} {\bibfield  {journal} {\bibinfo  {journal}
  {Phys. Rev. Lett.}\ }\textbf {\bibinfo {volume} {100}},\ \bibinfo {pages}
  {145301} (\bibinfo {year} {2008})}\BibitemShut {NoStop}%
\bibitem [{\citenamefont {Dusseault}\ and\ \citenamefont
  {Boninsegni}(2017)}]{marisa}%
  \BibitemOpen
  \bibfield  {author} {\bibinfo {author} {\bibfnamefont {M.}~\bibnamefont
  {Dusseault}}\ and\ \bibinfo {author} {\bibfnamefont {M.}~\bibnamefont
  {Boninsegni}},\ }\bibfield  {title} {\enquote {\bibinfo {title} {Atomic
  displacements in quantum crystals},}\ }\href {\doibase
  10.1103/PhysRevB.95.104518} {\bibfield  {journal} {\bibinfo  {journal} {Phys.
  Rev. B}\ }\textbf {\bibinfo {volume} {95}},\ \bibinfo {pages} {104518}
  (\bibinfo {year} {2017})}\BibitemShut {NoStop}%
\bibitem [{\citenamefont {Fernandez-Alonso}\ \emph {et~al.}(2012)\citenamefont
  {Fernandez-Alonso}, \citenamefont {Cabrillo}, \citenamefont
  {Fern\'andez-Perea}, \citenamefont {Bermejo}, \citenamefont {Gonz\'alez},
  \citenamefont {Mondelli},\ and\ \citenamefont {Farhi}}]{cabrillo}%
  \BibitemOpen
  \bibfield  {author} {\bibinfo {author} {\bibfnamefont {F.}~\bibnamefont
  {Fernandez-Alonso}}, \bibinfo {author} {\bibfnamefont {C.}~\bibnamefont
  {Cabrillo}}, \bibinfo {author} {\bibfnamefont {R.}~\bibnamefont
  {Fern\'andez-Perea}}, \bibinfo {author} {\bibfnamefont {F.~J.}\ \bibnamefont
  {Bermejo}}, \bibinfo {author} {\bibfnamefont {M.~A.}\ \bibnamefont
  {Gonz\'alez}}, \bibinfo {author} {\bibfnamefont {C.}~\bibnamefont
  {Mondelli}}, \ and\ \bibinfo {author} {\bibfnamefont {E.}~\bibnamefont
  {Farhi}},\ }\bibfield  {title} {\enquote {\bibinfo {title} {{Solid
  para-hydrogen as the paradigmatic quantum crystal: Three observables probed
  by ultrahigh-resolution neutron spectroscopy}},}\ }\href {\doibase
  10.1103/PhysRevB.86.144524} {\bibfield  {journal} {\bibinfo  {journal} {Phys.
  Rev. B}\ }\textbf {\bibinfo {volume} {86}},\ \bibinfo {pages} {144524}
  (\bibinfo {year} {2012})}\BibitemShut {NoStop}%
\bibitem [{\citenamefont {Mezzacapo}\ and\ \citenamefont
  {Boninsegni}(2009)}]{cqp}%
  \BibitemOpen
  \bibfield  {author} {\bibinfo {author} {\bibfnamefont {F.}~\bibnamefont
  {Mezzacapo}}\ and\ \bibinfo {author} {\bibfnamefont {M.}~\bibnamefont
  {Boninsegni}},\ }\bibfield  {title} {\enquote {\bibinfo {title} {Classical
  and quantum physics of hydrogen clusters},}\ }\href@noop {} {\bibfield
  {journal} {\bibinfo  {journal} {J. Phys.: Condens. Matter}\ }\textbf
  {\bibinfo {volume} {21}},\ \bibinfo {pages} {164205} (\bibinfo {year}
  {2009})}\BibitemShut {NoStop}%
\bibitem [{\citenamefont {Mezzacapo}\ and\ \citenamefont
  {Boninsegni}(2011)}]{supercl}%
  \BibitemOpen
  \bibfield  {author} {\bibinfo {author} {\bibfnamefont {F.}~\bibnamefont
  {Mezzacapo}}\ and\ \bibinfo {author} {\bibfnamefont {M.}~\bibnamefont
  {Boninsegni}},\ }\bibfield  {title} {\enquote {\bibinfo {title} {On the
  possible "supersolid" character of parahydrogen clusters},}\ }\href@noop {}
  {\bibfield  {journal} {\bibinfo  {journal} {J. Phys. Chem. A}\ }\textbf
  {\bibinfo {volume} {115}},\ \bibinfo {pages} {6831--6837} (\bibinfo {year}
  {2011})}\BibitemShut {NoStop}%
\bibitem [{\citenamefont {Sindzingre}\ \emph {et~al.}(1989)\citenamefont
  {Sindzingre}, \citenamefont {Klein},\ and\ \citenamefont
  {Ceperley}}]{sindzingre89}%
  \BibitemOpen
  \bibfield  {author} {\bibinfo {author} {\bibfnamefont {P.}~\bibnamefont
  {Sindzingre}}, \bibinfo {author} {\bibfnamefont {M.~L.}\ \bibnamefont
  {Klein}}, \ and\ \bibinfo {author} {\bibfnamefont {D.~M.}\ \bibnamefont
  {Ceperley}},\ }\bibfield  {title} {\enquote {\bibinfo {title} {{Path-integral
  Monte Carlo Study of Low-Temperature $^4$He Clusters}},}\ }\href {\doibase
  10.1103/PhysRevLett.63.1601} {\bibfield  {journal} {\bibinfo  {journal}
  {Phys. Rev. Lett.}\ }\textbf {\bibinfo {volume} {63}},\ \bibinfo {pages}
  {1601--1604} (\bibinfo {year} {1989})}\BibitemShut {NoStop}%
\bibitem [{\citenamefont {Kwon}\ \emph {et~al.}(2006)\citenamefont {Kwon},
  \citenamefont {Paesani},\ and\ \citenamefont {Whaley}}]{kwonls}%
  \BibitemOpen
  \bibfield  {author} {\bibinfo {author} {\bibfnamefont {Y.}~\bibnamefont
  {Kwon}}, \bibinfo {author} {\bibfnamefont {F.}~\bibnamefont {Paesani}}, \
  and\ \bibinfo {author} {\bibfnamefont {K.~B.}\ \bibnamefont {Whaley}},\
  }\bibfield  {title} {\enquote {\bibinfo {title} {Local superfluidity in
  inhomogeneous quantum fluids},}\ }\href@noop {} {\bibfield  {journal}
  {\bibinfo  {journal} {Phys. Rev. B}\ }\textbf {\bibinfo {volume} {74}},\
  \bibinfo {pages} {174522} (\bibinfo {year} {2006})}\BibitemShut {NoStop}%
\bibitem [{\citenamefont {Boninsegni}(2012)}]{su}%
  \BibitemOpen
  \bibfield  {author} {\bibinfo {author} {\bibfnamefont {M.}~\bibnamefont
  {Boninsegni}},\ }\bibfield  {title} {\enquote {\bibinfo {title} {{Supersolid
  phases of cold atom assemblies}},}\ }\href {\doibase
  10.1007/s10909-012-0571-1} {\bibfield  {journal} {\bibinfo  {journal} {J. Low
  Temp. Phys.}\ }\textbf {\bibinfo {volume} {168}},\ \bibinfo {pages}
  {137--149} (\bibinfo {year} {2012})}\BibitemShut {NoStop}%
\bibitem [{\citenamefont {Kora}\ and\ \citenamefont
  {Boninsegni}(2019)}]{patterned}%
  \BibitemOpen
  \bibfield  {author} {\bibinfo {author} {\bibfnamefont {Y.}~\bibnamefont
  {Kora}}\ and\ \bibinfo {author} {\bibfnamefont {M.}~\bibnamefont
  {Boninsegni}},\ }\bibfield  {title} {\enquote {\bibinfo {title} {{Patterned
  supersolids in dipolar Bose systems}},}\ }\href {\doibase
  10.1007/s10909-019-02229-z} {\bibfield  {journal} {\bibinfo  {journal} {J.
  Low Temp. Phys.}\ }\textbf {\bibinfo {volume} {197}},\ \bibinfo {pages}
  {337--347} (\bibinfo {year} {2019})}\BibitemShut {NoStop}%
\bibitem [{\citenamefont {Egan}(1981)}]{Egan}%
  \BibitemOpen
  \bibfield  {author} {\bibinfo {author} {\bibfnamefont {P.~O.}\ \bibnamefont
  {Egan}},\ }\bibfield  {title} {\enquote {\bibinfo {title} {{Muonic
  Helium}},}\ }in\ \href@noop {} {\emph {\bibinfo {booktitle} {{Atomic Physics
  7}}}},\ \bibinfo {editor} {edited by\ \bibinfo {editor} {\bibfnamefont
  {Daniel}\ \bibnamefont {Kleppner}}\ and\ \bibinfo {editor} {\bibfnamefont
  {Francis~M.}\ \bibnamefont {Pipkin}}}\ (\bibinfo  {publisher} {Plenum
  Press},\ \bibinfo {address} {New York},\ \bibinfo {year} {1981})\ pp.\
  \bibinfo {pages} {373--384}\BibitemShut {NoStop}%
\bibitem [{\citenamefont {Wheeler}(1988)}]{wheeler}%
  \BibitemOpen
  \bibfield  {author} {\bibinfo {author} {\bibfnamefont {J.~A.}\ \bibnamefont
  {Wheeler}},\ }\bibfield  {title} {\enquote {\bibinfo {title} {Nanosecond
  matter},}\ }in\ \href@noop {} {\emph {\bibinfo {booktitle} {Energy in
  Physics, War and Peace: A Festschrift Celebrating Edward Teller's 80th
  Birthday}}},\ \bibinfo {editor} {edited by\ \bibinfo {editor} {\bibfnamefont
  {H}~\bibnamefont {Mark}}\ and\ \bibinfo {editor} {\bibfnamefont
  {L.}~\bibnamefont {Wood}}}\ (\bibinfo  {publisher} {Kluwer},\ \bibinfo
  {address} {Dordrecht, The Netherlands},\ \bibinfo {year} {1988})\
  Chap.~\bibinfo {chapter} {10}, pp.\ \bibinfo {pages} {266--290}\BibitemShut
  {NoStop}%
\bibitem [{\citenamefont {Masaki}\ \emph {et~al.}(2020)\citenamefont {Masaki},
  \citenamefont {Aghai-Khozani}, \citenamefont {S\'ot\'er}, \citenamefont
  {Dax},\ and\ \citenamefont {Barna}}]{pionic}%
  \BibitemOpen
  \bibfield  {author} {\bibinfo {author} {\bibfnamefont {H.}~\bibnamefont
  {Masaki}}, \bibinfo {author} {\bibfnamefont {H.}~\bibnamefont
  {Aghai-Khozani}}, \bibinfo {author} {\bibfnamefont {A.}~\bibnamefont
  {S\'ot\'er}}, \bibinfo {author} {\bibfnamefont {A.}~\bibnamefont {Dax}}, \
  and\ \bibinfo {author} {\bibfnamefont {D.}~\bibnamefont {Barna}},\ }\bibfield
   {title} {\enquote {\bibinfo {title} {Laser spectroscopy of pionic helium
  atoms},}\ }\href@noop {} {\bibfield  {journal} {\bibinfo  {journal} {Nature}\
  }\textbf {\bibinfo {volume} {581}},\ \bibinfo {pages} {37--41} (\bibinfo
  {year} {2020})}\BibitemShut {NoStop}%
\bibitem [{\citenamefont {Fleming}\ \emph {et~al.}(1979)\citenamefont
  {Fleming}, \citenamefont {Garner}, \citenamefont {Vaz}, \citenamefont
  {Walker}, \citenamefont {Brewer},\ and\ \citenamefont {Crowe}}]{muonium}%
  \BibitemOpen
  \bibfield  {author} {\bibinfo {author} {\bibfnamefont {D.~G.}\ \bibnamefont
  {Fleming}}, \bibinfo {author} {\bibfnamefont {D.~M.}\ \bibnamefont {Garner}},
  \bibinfo {author} {\bibfnamefont {L.~C.}\ \bibnamefont {Vaz}}, \bibinfo
  {author} {\bibfnamefont {D.~C.}\ \bibnamefont {Walker}}, \bibinfo {author}
  {\bibfnamefont {J.~H.}\ \bibnamefont {Brewer}}, \ and\ \bibinfo {author}
  {\bibfnamefont {K.~M.}\ \bibnamefont {Crowe}},\ }\bibfield  {title} {\enquote
  {\bibinfo {title} {{Muonium Chemistry—A Review}},}\ }in\ \href {\doibase
  10.1021/ba-1979-0175.ch013} {\emph {\bibinfo {booktitle} {{Positronium and
  Muonium Chemistry. Advances in Chemistry Series}}}},\ Vol.\ \bibinfo {volume}
  {175},\ \bibinfo {editor} {edited by\ \bibinfo {editor} {\bibfnamefont
  {H.~J.}\ \bibnamefont {Ache}}}\ (\bibinfo  {publisher} {American Chemical
  Society},\ \bibinfo {address} {Washington},\ \bibinfo {year} {1979})\
  Chap.~\bibinfo {chapter} {13}, pp.\ \bibinfo {pages} {279--334}\BibitemShut
  {NoStop}%
\bibitem [{\citenamefont {Suffczy\'nski}\ \emph {et~al.}(2002)\citenamefont
  {Suffczy\'nski}, \citenamefont {Kotowski},\ and\ \citenamefont
  {Wolniewicz}}]{suff}%
  \BibitemOpen
  \bibfield  {author} {\bibinfo {author} {\bibfnamefont {M.}~\bibnamefont
  {Suffczy\'nski}}, \bibinfo {author} {\bibfnamefont {T.}~\bibnamefont
  {Kotowski}}, \ and\ \bibinfo {author} {\bibfnamefont {L.}~\bibnamefont
  {Wolniewicz}},\ }\bibfield  {title} {\enquote {\bibinfo {title} {{Size} of
  {Muonium} {Hydride}},}\ }\href {\doibase
  https://doi.org/10.12693/APhysPolA.102.351} {\bibfield  {journal} {\bibinfo
  {journal} {{Acta Phys. Pol. A}}\ }\textbf {\bibinfo {volume} {102}},\
  \bibinfo {pages} {351--354} (\bibinfo {year} {2002})}\BibitemShut {NoStop}%
\end{thebibliography}%






\end{document}